\documentclass[%
 reprint,
 amsmath,amssymb,
 aps,
superscriptaddress
]{revtex4-2}
\usepackage[utf8]{inputenc}
\usepackage{graphicx}
\usepackage{diagbox}
\usepackage[normalem]{ulem}
\usepackage[colorlinks,bookmarks=false,linkcolor=blue,urlcolor=blue]{hyperref}
\usepackage[top=2.5cm,bottom=2.5cm,left=2cm,right=2cm]{geometry}
\usepackage{color}
\usepackage[dvipsnames]{xcolor}
\usepackage{hyperref}
\usepackage{wrapfig}
\usepackage{multirow}
\usepackage{mathtools}
\usepackage{gensymb}
\usepackage{graphicx}
\usepackage{float}
\usepackage{bm}
\usepackage{upgreek}
\usepackage{textcomp}
\usepackage{booktabs}
\usepackage{units}
\usepackage{setspace}
\usepackage{subfiles}
\usepackage{color, colortbl}
\usepackage{makecell}
\definecolor{mygray}{gray}{0.6}
\definecolor{LightCyan}{rgb}{0.88,1,1}
\definecolor{Mycolor2}{rgb}{1,0.88,1}

\def \be {\begin{equation}}
\def \ee {\end{equation}}

\DeclarePairedDelimiter\abs{\lvert}{\rvert}%

\makeatletter
\let\oldabs\abs
\def\abs{\@ifstar{\oldabs}{\oldabs*}}
\makeatother

\date{\today}

\begin{document}

\title[Sample title]{
Influence of the  triangular Mn-O breathing mode on the magnetic ordering in multiferroic hexagonal manganites
}

\author{Tara N. Tošić}%
 \email{tara.tosic@mat.ethz.ch}
 
\affiliation{Materials Theory, ETH Zurich, Wolfgang-Pauli-Strasse 27, 8093 Z\"urich, Switzerland}%

\author{Quintin N. Meier}%

\affiliation{Université Grenoble Alpes, CEA, LITEN, 17 rue des Martyrs, 38054 Grenoble, France}
\affiliation{Materials Theory, ETH Zurich, Wolfgang-Pauli-Strasse 27, 8093 Z\"urich, Switzerland}%

\author{Nicola A. Spaldin}%

\affiliation{Materials Theory, ETH Zurich, Wolfgang-Pauli-Strasse 27, 8093 Z\"urich, Switzerland}%


\begin{abstract}
We use a combination of symmetry analysis, phenomenological modelling and first-principles density functional theory to explore the interplay
between the magnetic ground state and the detailed atomic structure in the hexagonal rare-earth manganites. We find that the magnetic ordering is sensitive to a breathing mode distortion of the Mn and O ions in the
$ab$ plane, which is described by the K\textsubscript{1} mode of the high-symmetry structure. Our density functional calculations of the
magnetic interactions indicate that this mode particularly affects the single-ion anisotropy and the inter-planar symmetric
exchanges. By extracting the parameters of a magnetic model Hamiltonian  from our first-principles results, we develop a phase diagram 
to describe the magnetic structure as a function of the anisotropy and exchange interactions. This in turn allows us to explain 
the dependence of the magnetic ground state on the identity of the rare-earth ion and on the K\textsubscript{1} mode.

\end{abstract}

\maketitle


\section{Introduction}
The hexagonal manganites, h-\textit{R}MnO$_3$, where \textit{R} = In, Sc, Y,  and Dy to Lu, are a class of multiferroic materials that show a combination of improper ferroelectricity and antiferromagnetism. Their hexagonal symmetry results in almost degenerate free energy surfaces in the hexagonal $ab$ plane for both the improper ferroelectric distortion \cite{artyukhin_landau_2014,meier_global_2017,skjaervo_unconventional_2019,meier_manifestion_2020} and for the magnetic order \cite{artyukhin_landau_2014,das_bulk_2014, marcela_magnetoelectric_2021}.
As a consequence of these quasi-degenerate ferroelectric and magnetic energy surfaces, small changes in the crystal chemistry lead to different structural and magnetic ground states. For example, in InMnO$_3$, small variations in the defect concentration favour either the improper ferroelectric state or a related antipolar phase \cite{kumagai_observation_2012, huang_duality_2014}. The magnetic energy surface in hexagonal manganites is even flatter. While all members of the series have a frustrated in-plane antiferromagnetic (AFM) arrangement of the Mn$^{3+}$ spin magnetic moments, the exact magnetic ground state varies with no obvious trend from compound to compound \cite{fiebig_determination_2000}. \par 
The goal of this work is to rationalize the evolution of the magnetic ground state across the hexagonal manganite series. We achieve this by decomposing the structural ground states into their distortions from the high-symmetry prototype structure, and determining the effects of these distortions on the magnetic interactions. Our main finding is that the  crystallographic K\textsubscript{1} mode, which consists of an in-plane triangular inwards or outwards breathing of the Mn and O ions, has a strong effect on the inter-planar exchanges and single-ion anisotropies, and ultimately determines the magnetic ground state of each material. Uncovering the details of this particular magneto-structural coupling sets the stage for engineering the magnetic order in this family of compounds.\par
\begin{figure*}
    \centering
    \includegraphics[scale=0.2175]{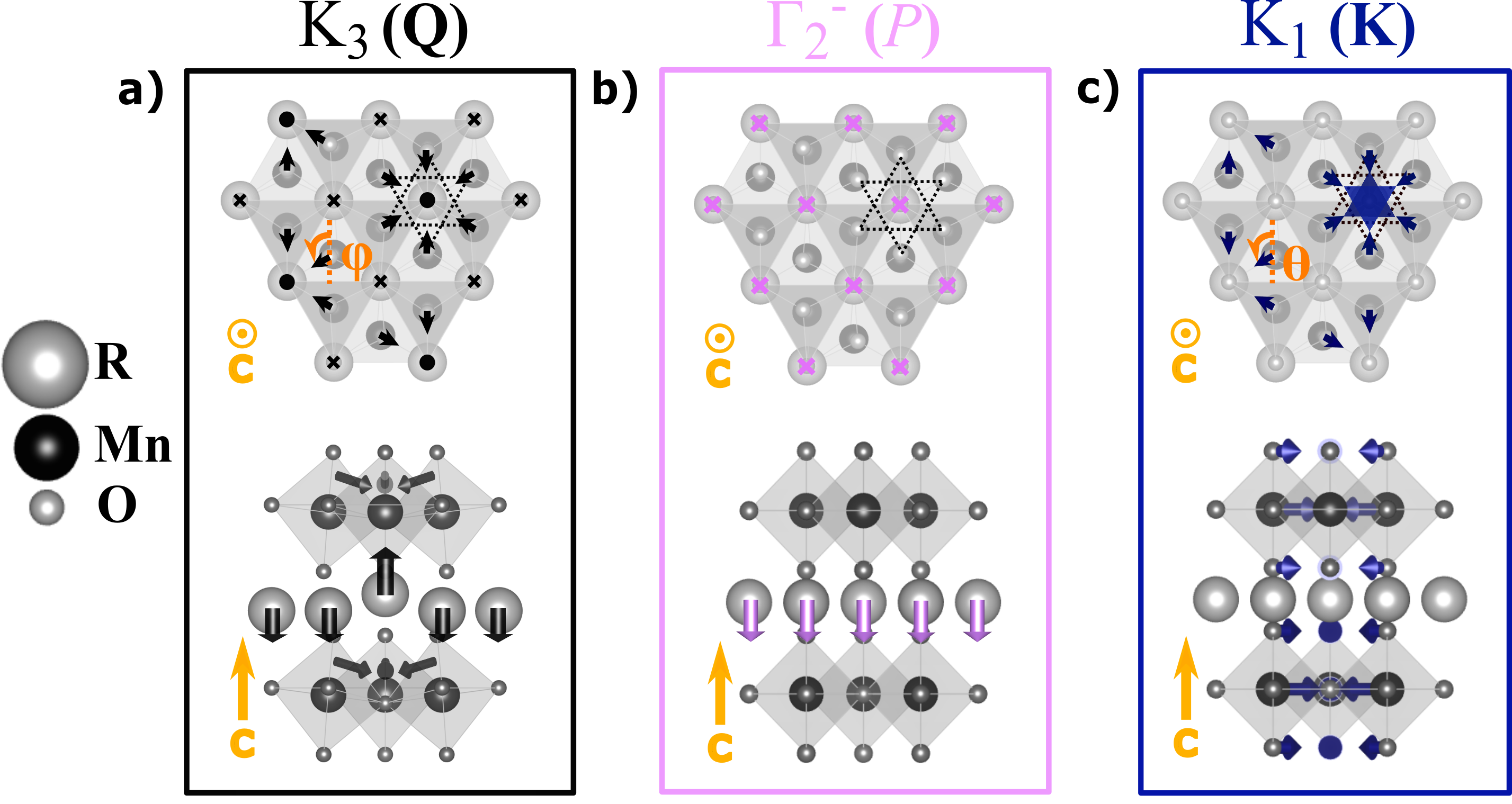}
    \caption{Structural distortions associated with the coupled K\textsubscript{3}, $\Gamma_2^-$ and K\textsubscript{1} modes and their respective order parameters \textbf{Q, P} and \textbf{K}. \textbf{a)} K\textsubscript{3}(\textbf{Q}). Black arrows represent the buckling of the \textit{R}  ions as well as the direction of the bipyramidal tilts, the latter characterised by the angle $\phi$ (in this example, $\phi=2\frac{\pi}{3}$). \textbf{b)} $\Gamma_2^-(\textbf{P})$. Pink arrows show the vertical displacement of the \textit{R}  ions associated with a negative polarisation along the $c$ axis. Note that displacements of O and Mn sites along $c$ also occur within $\Gamma_2^-$ but are not represented here. \textbf{c)} K\textsubscript{1}(\textbf{K}). Blue arrows indicate planar Mn and apical O displacements, their direction described by a characteristic angle $\theta$ (in the above example, $\theta=2\frac{\pi}{3}$). Shaded blue triangles show stacked triangular Mn trimers breathing in, relative to the high symmetry position Mn trimers (shown in dotted black lines), as a result of a negative $\delta x_{\text{Mn}}$ displacement.}
   \label{trimerisation}
\end{figure*}


\section{Structure}
The h-\textit{R}MnO$_3$ structure consists of layers of corner-sharing MnO$_5$ trigonal bipyramids alternating in the $c$ direction with triangular layers of \textit{R}-site cations. All members of the series adopt the non-polar P6$_3$/mmc space group at high temperature, and undergo a structural phase transition at a critical temperature T\textsubscript{c} that ranges from $1200-1600$\,K, depending on the radius of the \textit{R} site \cite{aken_hEMO_2001,bieringer_magneticstructure_1999}. At T\textsubscript{c}  the crystal symmetry lowers from non-polar P6$_3$/mmc to polar P6\textsubscript{3}cm, with the corresponding structural distortion consisting primarily of a K\textsubscript{3} mode, accompanied by a polar $\Gamma_2^-$ mode \cite{fennie_ferroelectric_2005}, of the high-symmetry structure. An additional smaller structural distortion, corresponding to a K\textsubscript{1} mode, can also occur, and has been associated with the onset of magnetic ordering at lower temperature \cite{howard_crystal_2013,lonkai_magnetic_2002}. We describe these distortion modes in detail next, reporting their contributions to both the experimentally observed structures and the Landau free energies which we extract using the INVARIANTS software from the group theoretical ISOTROPY Software Suite \cite{invariants1_stokes,invariants2_stokes_2003}.

\subsection{K\textsubscript{3} and $\Gamma_2^-$ modes  }

The primary order parameter driving the structural phase transition at T\textsubscript{c} describes a zone boundary mode at $\textbf{k}=(\nicefrac{1}{3},\nicefrac{1}{3},0)$, belonging to the K\textsubscript{3} irreducible representation of the high-symmetry P6$_3$/mmc structure  \cite{fennie_ferroelectric_2005,artyukhin_landau_2014,meier_global_2017,meier_manifestion_2020}. The distortion, illustrated in Fig.~\ref{trimerisation}\textbf{a)}, consists of triangles of MnO\textsubscript{5} trigonal bipyramids tilting towards or away from their corner-shared O site, accompanied by a buckling of the $R$-ion plane along the $c$ axis, and results in a trimerization of the lattice. We write the associated primary order parameter as $\textbf{Q}=\mathcal{Q}(\cos({\phi}),\sin(\phi))$, with $\mathcal{Q}$ giving the amplitude of the tilt and the angle $\phi$ its phase, as illustrated in Fig.~\ref{trimerisation}\textbf{a)}. The tilt angles have six-fold symmetry with $\phi=n\frac{\pi}{3}$ and the integer $n=1,...,6$. The ferroelectric polarisation $\mathcal{P}$ results from a net displacement of the \textit{R} ions along $c$ belonging to the $\Gamma_2^-$ irreducible representation of the P6$_3$/mmc structure. $\mathcal{P}$ emerges through a coupling to \textbf{Q}, established to be of the form \[f_{\textbf{Q}\mathcal{P}} \propto \mathcal{Q}^3\mathcal{P}\cos{(3\phi)}\] to lowest order in the Landau expansion of the free energy \cite{fennie_ferroelectric_2005,artyukhin_landau_2014, meier_global_2017}.

\subsection{K\textsubscript{1} mode and its coupling to K\textsubscript{3} and $\Gamma_2^-$}

In addition to the K\textsubscript{3} and the $\Gamma_2^-$ modes, a third structural distortion is reported at temperatures below T\textsubscript{c}, although with much smaller amplitude \cite{thomson_elastic_2014,toulouse_lattice_2014,lonkai_magnetic_2002,lee_giant_2008,chatterji_magnetoelastic_2012}; group theoretical analysis of the minimum energy structure of YMnO$_3$, calculated using density functional theory (DFT), indicated that its corresponding ionic displacements are one and two orders of magnitude smaller than those of the $\Gamma_2^-$ and K\textsubscript{3} modes, respectively \cite{fennie_ferroelectric_2005}. The mode belongs to the K\textsubscript{1} irreducible representation of the P6$_3$/mmc structure at the same  $\textbf{k}=(\nicefrac{1}{3},\nicefrac{1}{3},0)$ value as K\textsubscript{3}. K\textsubscript{1} involves collective planar displacements of the Mn ions and their apical O (O\textsubscript{ap}) ions parallel to the directions of the bipyramidal tilts projected onto the $ab$ plane, as depicted in Fig.~\ref{trimerisation}\textbf{c)}. Within this mode, the displacements of the O\textsubscript{ap} ions are much smaller than those of the Mn ions \cite{fennie_ferroelectric_2005}. Therefore, K\textsubscript{1} is discussed in terms of displacements $\delta x_{\text{Mn}}$ of the Mn ions away from their high symmetry positions at their Wyckoff site ($\nicefrac{1}{3}$,0,0) \cite{lonkai_magnetic_2002}. Throughout this work, $\delta x_{\text{Mn}}$ is expressed in units of fractional coordinates of the in-plane lattice parameters. Movements of Mn sites away ($\delta x_{\text{Mn}}>0$) or towards ($\delta x_{\text{Mn}}<0$) their corresponding trimerisation centres lead to a triangular breathing inwards and outwards of Mn sites that belong to the same trimer, as depicted by the respective contraction and expansion of blue coloured triangles in Fig.~\ref{trimerisation}\textbf{c)}.\par

\begin{table*}[tbp]
\setlength{\tabcolsep}{6pt}
    \centering
    \begin{tabular}{c c c c c c c}
    \toprule
    \toprule
        \multirow{2}{*}{}  & Sc & Lu & Yb &Er & Ho&Y\\[+5pt]
        Radius [\AA] & 0.87&0.977 & 0.985& 1.004& 1.015& 1.019\\
        \midrule
        
          $T\geq T\textsubscript{N}$ & 
          \makecell[c]{-0.0001(72)$^{\ddagger}$ \cite{munoz_magnetic_2000}\\ \hspace{3pt}0.0008(85)$^{\ddagger}$ \cite{fabreges_spin-lattice_2009}}&
          -0.001$^{\ddagger}$ \cite{lee_giant_2008} &
          -0.0063(18)$^{\ddagger}$ \cite{fabreges_spin-lattice_2009}& 
          \makecell{-0.020$^{\dagger}$ \cite{liu_hEMO_2011}\\-0.005$^{\dagger}$  \cite{aken_hEMO_2001}}& 
          \makecell[c]{-0.0072(12)$^{\ddagger}$ \cite{fabreges_spin-lattice_2009}\\\hspace{-7pt}0.002(8)$^{\ddagger}$ \cite{lonkai_magnetic_2002}}& 
          \makecell[l]{
              -0.0155(4)$^{\ddagger}$ \cite{gibbs_high-temperature_2011}\\
              -0.0125(2)$^{\ddagger}$ \cite{munoz_magnetic_2000}\\
              -0.00861$^{\ddagger}$ \cite{singh_dominance_2010}\\
              -0.0003(16)$^{\ddagger}$ \cite{lee_giant_2008}\\ 
              \hspace{3pt}0.004(8)$^{\ddagger}$$^{*}$ \cite{lonkai_magnetic_2002}} \\

        \midrule
        
         $T\leq 10$\,K & 
         
         -0.0029(16)$^{\ddagger}$ \cite{fabreges_spin-lattice_2009}&
         -0.003$^{\ddagger}$ \cite{lee_giant_2008} & 
         -0.0023(19)$^{\ddagger}$ \cite{fabreges_spin-lattice_2009}&
         (-) & 
         \makecell{\hspace{-10pt}-0.003(2)$^{\ddagger}$ \cite{lonkai_magnetic_2002}\\
         \hspace{3pt}0.0025(86)$^{\ddagger}$ \cite{fabreges_spin-lattice_2009}}&
         \makecell{0.0001(7)$^{\dagger}$  \cite{brown_neutron_2006}\\\hspace{-9pt}0.0089$^{\ddagger}$ \cite{lee_giant_2008}}\\
         
         \bottomrule
         \bottomrule
         
    \end{tabular}
    \caption{Diffraction measurements of $\delta x_{\text{Mn}}$ displacements in the hexagonal manganites above (upper panel) and below (lower panel) the N\'{e}el temperature, $T\textsubscript{N}$. The Shannon radii (for eight-coordinated 3+ states) are shown to indicate the radial trend \cite{Shannon_revised_1976}. $^{\ddagger}$Powder sample. $^{\dagger}$Single crystal sample. $^{*}$Sample contained oxygen deficiency of 0.29(3) per formula unit. (-) No data available.}
    \label{tab:deltax_inhRMO}
\end{table*}

The K\textsubscript{1} mode can also be described within the Landau free energy expansion by its order parameter $\textbf{K}=\mathcal{K}\left(\cos{(\theta)};\sin{(\theta)}\right)$, where $\theta$ describes the direction of the Mn and apical O displacements and $\mathcal{K}$ is their amplitude. The K\textsubscript{1} mode is like the $\Gamma_2^-$ mode in that it is stable in the high symmetry structure and it differs in that it is constrained by a three fold symmetry; i.e.

\begin{equation}
    f_{\mathcal{K}}=\beta_1\mathcal{K}^2+\beta_2\mathcal{K}^3\cos{(3\theta)}
\end{equation}

Since K\textsubscript{3} is the primary order parameter \cite{fennie_ferroelectric_2005}, the two coupling strengths $\beta_1$ and $\beta_2$ are positive and K\textsubscript{1} emerges through its couplings to the K\textsubscript{3} and $\Gamma_2^-$ modes. These are third-order terms in the Landau free energy, with the form:

\begin{equation}\label{PK1K3_LFE}
    f_{\textbf{KQ}\mathcal{P}}=\gamma_1\mathcal{KQ}^2\cos{(\theta + 2\phi)}+\gamma_2 \mathcal{PKQ}\cos{(\theta-\phi)}\quad ,
\end{equation}

that is linear quadratic between K\textsubscript{1} and K\textsubscript{3} and trilinear between $\Gamma_2^-$, K\textsubscript{1} and K\textsubscript{3}. K\textsubscript{1} emerges improperly with $\gamma_{1,2}<0$. Given the six possible values of $\phi$ and that $\mathcal{P}$ alternates sign between consecutive values of $n$ \cite{artyukhin_landau_2014}, Eq.~\eqref{PK1K3_LFE} is minimised for $\theta=\pi -2\phi$ or $\theta=-2\phi$. Thus, the movement of the Mn and apical O sites within the K\textsubscript{1} mode is restricted along the direction defined by $\phi$ as illustrated in Fig.~\ref{trimerisation}.\par

In Table \ref{tab:deltax_inhRMO} we list reported measured room-temperature and sub-T\textsubscript{N} values of $\delta x_{\text{Mn}}$ for six hexagonal manganites. The data illustrate three points.  First, $\delta x_{\text{Mn}}$ is small and therefore difficult to quantify, reflected in a spread of reported values. Second, there is no obvious trend in $\delta x_{\text{Mn}}$ across the series. Finally, studies that measure Mn positions above and below $T\textsubscript{N}$ report trimers of Mn sites expanding below T\textsubscript{N} for YbMnO$_3$ and YMnO$_3$ but contracting for ScMnO$_3$ and LuMnO$_3$, with no clear difference for HoMnO$_3$. Low temperature measurements have not been made for ErMnO$_3$. These observations lead to three open questions, which we address in this work: First, is $\delta x_{\text{Mn}}$ non zero? Second, is there a trend in $\delta x_{\text{Mn}}$ across the series? Third, what is the mechanism behind any activation of the K\textsubscript{1} mode, and is it temperature-dependent \cite{lee_giant_2008} or temperature-independent \cite{lonkai_magnetic_2002}?


\section{Magneto-structural coupling}

Having described the relevant structural distortions, we now turn to the magnetic properties and thus to the main objective of this study: the coupling between the magnetic order and the crystallographic structure.

\subsection{Magnetic symmetry and properties}

The small trigonal-bipyramidal crystal field splitting combined with exchange interaction favour the high-spin state on the formally $d^4$, L=0 Mn$^{3+}$ ions \cite{das_bulk_2014}.  These Mn$^{3+}$ magnetic moments order at a Néel temperature, T\textsubscript{N} $\simeq 70\-- 90$\,K, with higher T\textsubscript{N} values corresponding to smaller \textit{R}-site radii \cite{thomson_elastic_2014,gibbs_high-temperature_2011,lorenz_hexagonal_2013}. The first nearest-neighbour interaction is strongly AFM and geometrically frustrated because of the triangular arrangement of the Mn ions \cite{fiebig_determination_2000}.\par

\begin{figure}[h!]
    \centering
    \includegraphics[scale=0.15]{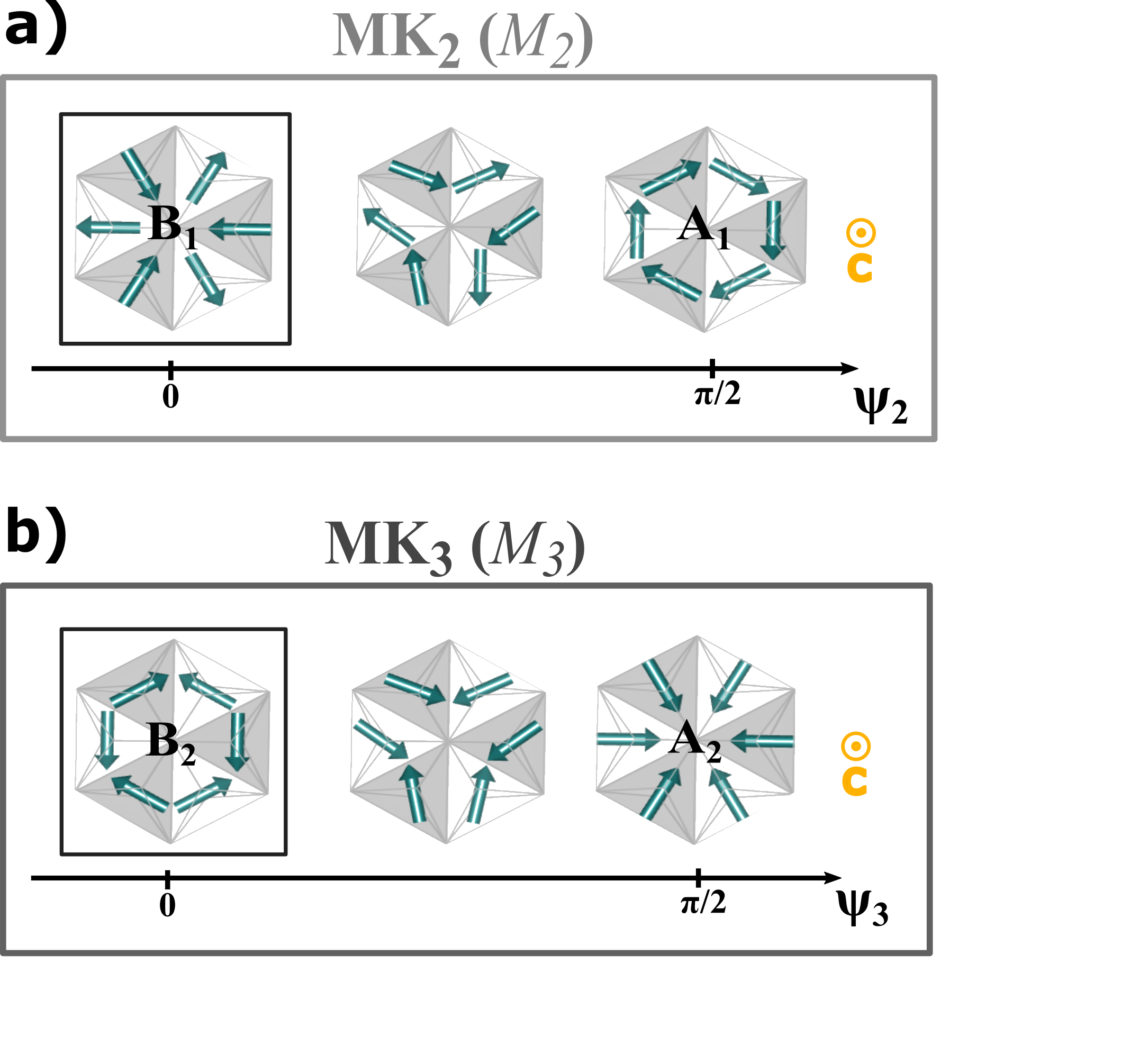}
    \caption{Planar magnetic configurations associated with the two magnetic modes mK\textsubscript{2} and mK\textsubscript{3}, corresponding to order parameters $\mathcal{M}_2$ and $\mathcal{M}_3$. The magnetic moments on each Mn site are represented by teal arrows. Consecutive planes are coloured in different shades. \textbf{a)} $\psi_2$ describes the evolution from B\textsubscript{1} ($\psi_2=0$) to A\textsubscript{1} ($\psi_2=\frac{\pi}{2}$) via an out-of-phase rotation of spins belonging to two consecutive planes. \textbf{b)} $\psi_3$ describes the evolution from B\textsubscript{2} ($\psi_3=0$) to A\textsubscript{2} ($\psi_3=\frac{\pi}{2}$) via an in-phase rotation of spins belonging to two consecutive planes. }
    \label{fig:LFE_K1_mKi_coupling}
\end{figure}

There are four symmetry-allowed candidate magnetic irreducible representations (irreps), labelled A\textsubscript{1,2} and B\textsubscript{1,2}, all of which have $120\degree$ first nearest-neighbour in-plane configurations. These irreps are generated under the zone boundary magnetic modes mK\textsubscript{2} or mK\textsubscript{3} described by the following respective order parameters: \textbf{M}\textsubscript{2}=$\mathcal{M}_2\left(\cos(\psi_2);\sin{(\psi_2)}\right)$ and \textbf{M}\textsubscript{3}=$\mathcal{M}_3\left(\cos(\psi_3);\sin{(\psi_3)}\right)$ as shown in Fig.~\ref{fig:LFE_K1_mKi_coupling} \cite{howard_crystal_2013}. The amplitudes $\mathcal{M}$\textsubscript{2,3} reflect the amount of correlation of magnetic moments on symmetry equivalent magnetic sites, and the angles $\psi_{\text{2,3}}$ describe the local direction of the Mn magnetic moments as sketched in Figs.~\ref{fig:LFE_K1_mKi_coupling}\textbf{a)} and \ref{fig:LFE_K1_mKi_coupling}\textbf{b)}. Within the four irreps, magnetic moments order either radially (A\textsubscript{2} and B\textsubscript{1}) or tangentially (A\textsubscript{1} and B\textsubscript{2}). The radial irreps A\textsubscript{2}  and B\textsubscript{1} also allow for a weak out-of-plane ferromagnetic (FM) and AFM canting, respectively \cite{howard_crystal_2013, das_bulk_2014}. Spins belonging to consecutive layers along the $c$ axis can order with even (A) or odd (B) symmetry under the two-fold screw rotation $2\Tilde{c}$ \cite{artyukhin_landau_2014}. This results in two magnetic moments aligned along the same axis but belonging to consecutive planes pointing either parallel or anti-parallel to each other, corresponding respectively to a B- or A-type ordering. All the orders generated under mK\textsubscript{2} and mK\textsubscript{3} would have the same magnetic energy in the high symmetry structure. This energy degeneracy is broken as the structural symmetry is lowered to P6\textsubscript{3}cm.

\subsection{Coupling between K\textsubscript{3} and mK\textsubscript{2,3} modes}

\begin{figure}[h!]
    \centering
    \includegraphics[scale=0.33]{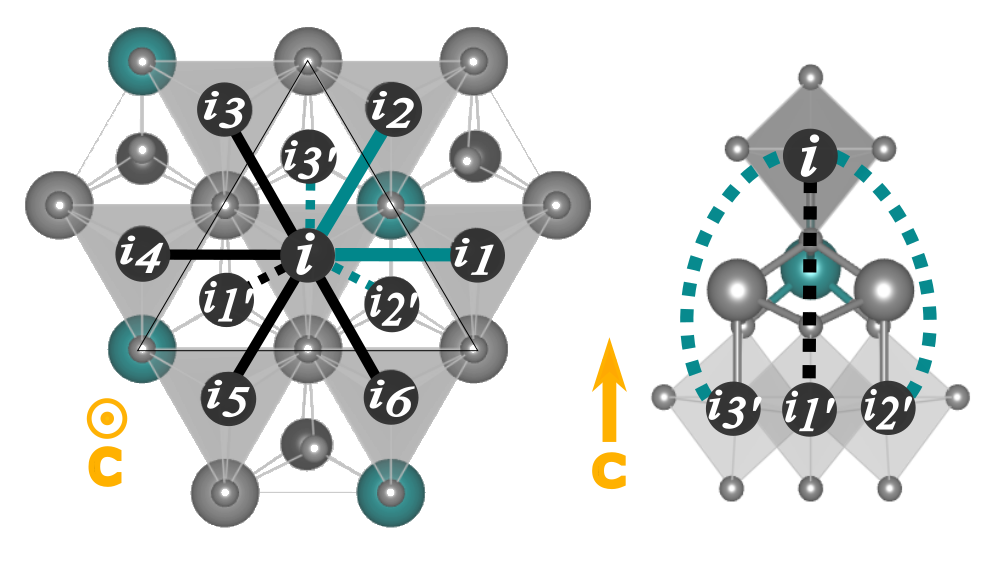}
    \caption{Symmetric exchanges of a magnetic site \textit{\textbf{i}} in the low symmetry P6\textsubscript{3}cm structure. Teal coloured \textit{R} sites indicate trimerisation centres; nearest-neighbour sites $i_1$ and $i_2$ occupy the same trimer as site $i$, whereas sites $i_3$, $i_4$, $i_5$ and $i_6$ occupy  different trimers. Solid teal and black lines represent same trimer ($J_{st}$) and  different trimer ($J_{dt}$) nearest-neighbour exchanges, respectively. Dashed teal and black lines represent the inter-planar $J_{1z}$ (with site $i_{1'}$) and $J_{2z}$ (with sites $i_{2'}$ and $i_{3'}$).}
    \label{symm_exch}
\end{figure}

 Magnetic order sets in on the low symmetry structure and this is expressed by a coupling between the primary order parameter and the magnetic order at fourth order in the Landau free energy expansion of the form \cite{das_bulk_2014,marcela_magnetoelectric_2021,artyukhin_landau_2014,fiebig_observation_2002} 
 \[f_{\textbf{Q,M}_{2,3}}\propto\mathcal{M}^2_{2,3}\mathcal{Q}^2\cos^2{(2\phi-2\psi_{2,3})}\quad .\] 
 As a consequence of this coupling, there are two types of in-plane nearest-neighbour interactions: those between two Mn ions that share a trimerization center ($J_{same\text{ }trimer\text{ }(st)}$) (solid teal lines in Fig.~\ref{symm_exch}), and those occupying neighbouring trimerization centers ($J_{different\text{ }trimer\text{ }(dt)}$) (solid black lines in Fig.~\ref{symm_exch}). The low symmetry P6\textsubscript{3}cm structure also has two different inter-planar second nearest-neighbour exchanges, $J_{1z}$ and $J_{2z}$. A total of six second nearest-neighbours interact with each given site $i$ either via $J_{1z}$ or $J_{2z}$, the former mediated by two equivalent (both grey in Fig.~\ref{symm_exch}) \textit{R} sites and the latter by two non-equivalent (one grey and one teal in Fig.~\ref{symm_exch}) \textit{R} sites. Finally, trigonal bipyramids have easy-plane single-ion anisotropy (SIA), with the hard axis tilted away from the $c$ direction by the K\textsubscript{3} mode. \\

Measurement of the exact magnetic ground state in the hexagonal manganites via scattering techniques is complicated by the issue of homometry \cite{howard_crystal_2013}; magnetic symmetries obtained within one of the mK\textsubscript{i=1,2} modes lead to near equality of their magnetic structure factors in scattering data if $\delta x_{\text{Mn}}=0$ \cite{howard_crystal_2013,chatterji_magnetoelastic_2012}. This is because the high symmetry position of the Mn sites (at Wyckoff position ($x\simeq\frac{1}{3}$,0,0), corresponding to $\delta x_{\text{Mn}}\simeq0$) lies in a mirror plane, $m_{\perp}[120]$ \cite{lonkai_magnetic_2002}. One way to address this issue is by using polarised neutron scattering \cite{brown_neutron_2006}, since the polarised character of the incident beam allows the different $\psi_2$ and $\psi_3$ values to be distinguished, or second harmonic generation techniques \cite{fiebig_determination_2000} which are directly sensitive to the symmetry.  B\textsubscript{1} and B\textsubscript{2} configurations (framed in black in Fig.~\ref{fig:LFE_K1_mKi_coupling}) have been observed for different members of the series using optical second harmonic spectroscopy, with no evidence of A-type order \cite{fiebig_determination_2000}.

\subsection{Coupling between K\textsubscript{1} and mK\textsubscript{2,3} modes}

Next, we review the experimental evidence for a sub-T\textsubscript{N} $\delta x_{\text{Mn}}$ structural distortion that motivates our investigation of a K\textsubscript{1}-mK\textsubscript{2,3} coupling \cite{chatterji_magnetoelastic_2012,thomson_elastic_2014,lee_giant_2008}. In YMnO\textsubscript{3}, powder diffraction measurements indicate that the experimental crystal volume deviates from the Einstein-Grüneisen volume predicted for a non-magnetic system \cite{chatterji_magnetoelastic_2012,sharma_spin_2013}. Additionally, resonant ultrasound spectroscopy detects an elastic stiffening in YMnO\textsubscript{3} as the crystal is cooled through T\textsubscript{N} \cite{thomson_elastic_2014}. Both experimental techniques show that lattice strain scales with the square of the magnetic order parameter \cite{chatterji_magnetoelastic_2012,thomson_elastic_2014}. The measured onset of this strain slightly above T\textsubscript{N} \cite{thomson_elastic_2014} alludes to a magneto-structural coupling that coincides with the emergence of short range correlations between spins.

\begin{figure}[h!]
    \centering
    \includegraphics[scale=0.15]{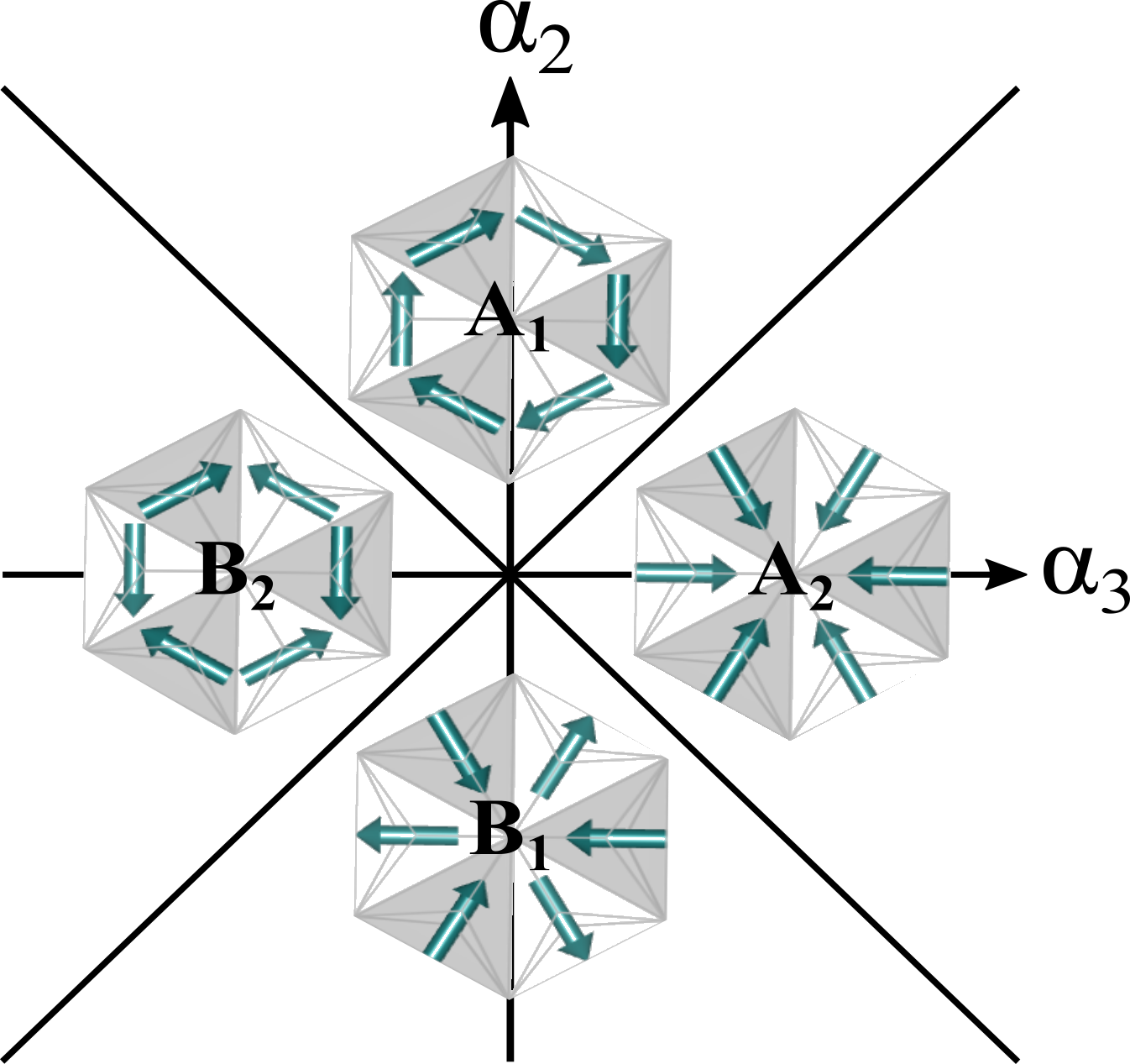}
    \caption{Magnetic ground state in the phase space of the K\textsubscript{1}-mK\textsubscript{2,3} coupling parameters $\alpha_{2,3}$. Diagonal lines indicate $|\alpha_2|=|\alpha_3|$.}
    \label{fig:landau_para_gsdiagram}
\end{figure}

We will now motivate the K\textsubscript{1}-mK\textsubscript{2,3} coupling using symmetry arguments. The lowest order magneto-structural coupling term that appears in the Landau Free Energy expansion and which remains invariant under the symmetry operations of the P6\textsubscript{3}/mmc phase is at third order between the K\textsubscript{1} and the mK\textsubscript{2,3} modes:
\begin{equation}\label{K1M_LFE}
\begin{aligned}
f_{\textbf{K,M}_{2,3}}&=\alpha_2\mathcal{KM}_{2}^2\cos{(\theta+2\psi_{2})} \\
&+ \alpha_3\mathcal{KM}_{3}^2\cos{(\theta+2\psi_{3})} \quad,
\end{aligned}
\end{equation}

where $\alpha_{2,3}$ are the coupling strengths. For a given K\textsubscript{1} displacement direction $\theta$, these coupling terms are minimised for $i=2,3$, by solving:

\begin{equation}
\begin{aligned}\label{LFE_minimise}
    &\left.\frac{\partial f_{\textbf{K,M}_{2,3}}}{\partial\psi_i}\right|_{\theta,\phi}=0\\
    &\Rightarrow-2\alpha_i\mathcal{KM}_i^2\sin{(\theta+2\psi_i)}=0\quad .\\
\end{aligned}
\end{equation}

Using Eq.~\eqref{LFE_minimise}, we calculate the ground-state magnetic phase diagram in the phase space of the two K\textsubscript{1}-mK\textsubscript{2} and K\textsubscript{1}-mK\textsubscript{3} coupling parameters ($\alpha_2$ and $\alpha_3$), shown in Fig.~\ref{fig:landau_para_gsdiagram}.

\subsection{Model magnetic Hamiltonian}

Next, in order to gain a microscopic understanding of the dependence of the magnetic ordering on the K\textsubscript{1} mode, we write down a model Hamiltonian, and calculate the interaction parameters as a function of the size and orientation of \textbf{K}. We describe the energy of the magnetic moment on site $i$ using the following Hamiltonian:

 \begin{align}\label{1stNN_AFM}
     \mathcal{H}_i=&\underbrace{\hat{e}_i\cdot\left(J_{st}\sum_{j=1,2}\hat{e}_{i_j}+J_{dt}\sum_{\substack{j=3,4, \\       5,6}}\hat{e}_{i_j}\right)}_{\text{1\textsuperscript{st} nearest-neighbour exchange}}\\  \label{2ndNN}
     &+\underbrace{\hat{e}_i\cdot2\left(J_{1z}\hat{e}_{i_{1'}} + J_{2z}\sum_{j'=2',3'}\hat{e}_{i_{j'}}\right)}_{\text{2\textsuperscript{nd} nearest-neighbour exchange}} \\\label{sia}
     &+\underbrace{A\cos{(\psi^i_{2,3})}}_{\text{SIA}} \quad .
 \end{align}

Here, $\hat{e}_i=(e_i^x,e_i^y,e_i^z)$ is the normalised magnetic moment of a site $i$ where $i=1$,...,6 designates one of the six magnetic sites of the unit cell and the indices $i_j$ and $i_{j'}$ run over the first and second nearest-neighbours of site $i$, respectively, as sketched in Fig.~\ref{symm_exch}. The nearest-neighbour couplings (mediated by the $J_{st}$ and $J_{dt}$ interactions) contribute equally to the energy in all four magnetic configurations, as can be seen by the $120\degree$ arrangement between nearest-neighbour sites in Fig.~\ref{fig:LFE_K1_mKi_coupling}. The difference in energy between the A- and B-type configurations results from the inter-planar ordering and is described by the second term of $\mathcal{H}_i$: $J_{1z}$, where $J_{2z}$ are the inter-planar exchanges illustrated in Fig.~\ref{fig:LFE_K1_mKi_coupling}. The third term \eqref{sia} gives the in-plane SIA, where $A$ is defined as the energy difference between a spin pointing towards its trimerisation centre and perpendicular to it; $\psi^i_{2,3}$ is the angle describing the magnetic order within the two magnetic modes mK\textsubscript{2,3} following the description in Figs.~\ref{fig:LFE_K1_mKi_coupling}\textbf{a)} and \ref{fig:LFE_K1_mKi_coupling}\textbf{b)}. Since the out-of-plane canting is known to be small we neglect any contribution to the total energy from rotation into the hard axis (see Methods section for further details).\par

 Taking advantage of the fact that on a frustrated triangular lattice $\hat{e}_{i2'}+\hat{e}_{i3'}=-\hat{e}_{i1'}$, the interactions lifting the magnetic energy degeneracy between A$_1$, A$_2$, B$_1$ and B$_2$ can be reduced to an  effective inter-planar exchange term $J_{z}=J_{1z}-J_{2z}$ and an effective in-plane anisotropy term $A$. Thus, we obtain the following single-spin Hamiltonian, reduced to only two relevant interactions, $J_{z}$ and $A$, in which the first-nearest-neighbour AFM contributions have been absorbed into $\mathcal{H}_{0}$:

\begin{align}
    \mathcal{H}_i&=\hat{e}_i\cdot2\left[J_{1z}\hat{e}_{i1'} + J_{2z}(\underbrace{\hat{e}_{i2'}+\hat{e}_{i3'}}_{=-\hat{e}_{i1'}})\right]\\
    \nonumber
    &\quad+ A\cos{(\psi^i_{2,3})} + \mathcal{H}_0 \\ 
    \nonumber\\ \label{Hamiltonian_eq}
    &=2J_z\hat{e}_i\cdot\hat{e}_{i1'}+A\cos{(\psi^i_{2,3})} + \mathcal{H}_0 \quad .
\end{align}

Within this model, the energy, $E_i$, of each spin $\hat{e}_i$ for each of the four allowed in-plane magnetic configurations is given by:

\begin{align}
    E_i(A_1)&= - 2J_{z} + \mathcal{H}_0 \quad, \\
    E_i(A_2)&= - 2J_{z} + A + \mathcal{H}_0\quad, \\
    E_i(B_1)&= 2J_{z} + A + \mathcal{H}_0 \quad \text{and} \\
    E_i(B_2)&= 2J_{z} + \mathcal{H}_0 \quad .
\end{align}


\section{Methods}
Total energies are obtained within DFT \cite{anisimov_LdaU_1997} based on the projector augmented-wave method \cite{Blochl_PAW_1994} as implemented in the Vienna Ab initio Simulation Package (VASP 5.4.4) \cite{vasp1,vasp2,vasp3,vasp4}. Calculations are performed using the Perdew-Burke-Ernzerhof (PBE) generalised gradient approximation \cite{perdew_generalized_1996,pbe_vasp} combined with an on-site Coulomb repulsion of U=4\,eV \cite{das_bulk_2014} and an exchange parameter of J=1\,eV (following the Liechtenstein approach \cite{liechtenstein_density-functional_1995}) on the Mn sites \cite{wang_structural_2014,das_bulk_2014}. We use the Y\_sv, Er\_3, Lu, Mn and O VASP library pseudopotentials, with $4d$, $5p$, $5p$ and $p$ electrons in the valence band, respectively. We compare the computed E(B\textsubscript{2})-E(B\textsubscript{1}) energy difference with calculated values obtained via a different approximation, the local density density approximation, and find that the energy trend stays the same, albeit with an energy difference of $\simeq0.2$\,meV. The effects of the U and J values on the computed energies are also evaluated: there is no notable difference between U=4 and U=6 whilst E(B\textsubscript{2})-E(B\textsubscript{1}) presents a steeper energy trend for J=1 than for J=0 with a maximal difference at big negative $\delta x_{\text{Mn}}$ displacements of the order of 0.1\,meV. The trends thus seem robust to our choice of U and J as well as our choice of pseudopotential. Finally, we use a cut-off of 700\,eV and a gamma-centred k-point grid of $6\times 6\times 3$.\par

\begin{figure*}
    \centering
    \includegraphics[scale=0.3]{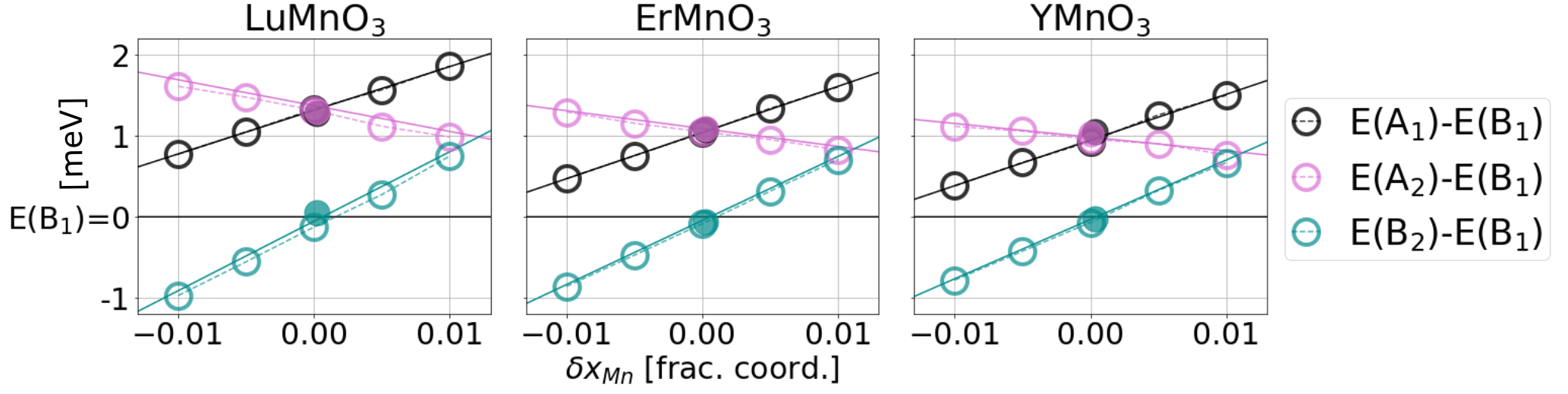}
    \caption{Total energies of the planar magnetic configurations A\textsubscript{1}, A\textsubscript{2} and B\textsubscript{2}, relative to the energy of B$_1$. Empty circles correspond to DFT calculated total energies for $\delta x_{\text{Mn}}\in[-0.01,0.01]$ fractional coordinates and the dotted lines describe their energy trend. Full lines represent the model Hamiltonian calculated energies. Full circles are DFT total energies calculated on fully relaxed geometries.}
   \label{model_vs_DFT_E_YELMO}
\end{figure*}
Geometries corresponding to the different values of the K\textsubscript{1} mode are obtained by selective dynamics relaxation of the cell, in which the Mn ions are fixed at positions away from their K\textsubscript{1}=0 high symmetry position and the other atomic positions as well as lattice parameters are relaxed. Within selective dynamics, the $\frac{c}{a}$ ratio decreases quadratically and symmetrically around $\delta x_{\text{Mn}}=0$ as Mn ions breathe inwards or outwards. This lattice ratio decreases by $\simeq0.002$ for the computed compounds as Mn ions shift by $0.01\times a$ from their high symmetry position. Note that, although K\textsubscript{1} is the only mode that allows a planar movement of Mn ions, the control of K\textsubscript{1} by a constraint on Mn positions within selective dynamics is not perfect. For example, the $\delta x_{\text{Mn}}=0$ geometry still allows for very small displacements of O\textsubscript{ap} within K\textsubscript{1}, and $\Gamma_2^-$ and K\textsubscript{3} are activated, as they couple to K\textsubscript{1}, when the amplitude of \textbf{K} is increased. However, the effect of these couplings is relatively small and will be the object of future works. For simplicity, we use the terms K\textsubscript{1} amplitude and $\delta x_{\text{Mn}}$ interchangeably in this work. Important to note is that, even though Mn ions also relax to $\delta x_{\text{Mn}}\simeq0$ in the fully relaxed geometries, the latter and the selective dynamics $\delta x_{\text{Mn}}=0$ structures differ across the three compounds. The fully relaxed geometries have higher $\frac{c}{a}$ ratios, relative to their selective dynamics relaxed counterparts at $\delta x_{\text{Mn}}=0$; namely 1.8788, 1.8654 and 1.8726 compared to 1.8900, 1.8654 and 1.8579 for LuMnO\textsubscript{3}, ErMnO\textsubscript{3} and YMnO\textsubscript{3}, respectively. Compared to the $\delta x_{\text{Mn}}=0$ structure, the fully relaxed geometries have higher displacements corresponding $\Gamma_1^+$ and K\textsubscript{3} (by $\simeq0.01$ and $\simeq 0.015$\,\r{A} respectively), as well as lower $\Gamma_2^-$ displacements (by $\simeq 0.002$\,\r{A}). However, the K\textsubscript{1} mode amplitude is similarly small in both geometries. These structural differences are important to keep in mind when comparing the results for the fully relaxed and $\delta x_{\text{Mn}}=0$ geometries.\par
A complete description of the SIA can be expressed through the $\tau_i$ matrix that is of the following form in each magnetic site's local coordinate system (defined for each Mn site with the local $x$ axis pointing towards the trimerisation centre):

\begin{equation}
    \tau_i=\begin{pmatrix}
    A_{xx}&0&A_{xz}\simeq0\\
    0&A_{yy}&0\\
    A_{xz}\simeq0&0&A_{zz}
    \end{pmatrix}
\end{equation}

We find that the weak out-of-plane canting that is allowed in the A\textsubscript{2} and B\textsubscript{1} configurations \cite{das_bulk_2014} contributes negligibly to the total magnetic energy: the computed values border on the limit of the numerical precision of our DFT computations ($10^{-5}$\,eV). As a consequence, we do not include them in our model Hamiltonian.\par
Next, we describe how we compute the Hamiltonian parameters in Eq.~\eqref{Hamiltonian_eq}. From Eqs.~\eqref{A1_eq} and \eqref{A2_eq}, we see that the $J_z$ values can be obtained by subtracting the energy of configuration A\textsubscript{2} (A\textsubscript{1}) from that of B\textsubscript{1} (B\textsubscript{2}). We take the $J_z$ value that is the average of the result from the two calculations (as shown in Eq.~\eqref{Jz_av}): 
 
\begin{align}
\label{Jz1_eq}
    J_z &= \frac{1}{4}(E_i(B_1) - E_i(A_2)) \quad ,\\
\label{Jz2_eq}
    &= \frac{1}{4}(E_i(B_2) - E_i(A_1))\\
\label{Jz_av}
    &\quad\Rightarrow J_z=\frac{\eqref{Jz1_eq}+\eqref{Jz2_eq}}{2}  \quad . 
\end{align}    

Similarly, the $A$ parameters are obtained by subtracting the energy of the A\textsubscript{1} (B\textsubscript{1}) configuration from that of A\textsubscript{2} (B\textsubscript{2}), and we take the average:
    
\begin{align}
\label{A1_eq}
    A &= E_i(A_2) - E_i(A_1) \quad ,\\
\label{A2_eq}
    &= E_i(B_1) - E_i(B_2) \\
\label{A_av}
    &\quad\Rightarrow A=\frac{\eqref{A1_eq}+\eqref{A2_eq}}{2}
    \quad . 
\end{align}

We then re-calculate $A$ and $J_z$ for a range of $\delta x_{\text{Mn}}$ values and make a linear fit of their dependence on $\delta x_{\text{Mn}}$ using a least squares method. The magnetic energies as a function of K\textsubscript{1} for different magnetic orders are obtained by substituting the $A$ and $J_z$ parameters into the Hamiltonian of Eq.~\eqref{Hamiltonian_eq}. We verify the accuracy of our model and the extracted parameters by comparing our model magnetic energy trends following Eq.~\eqref{Hamiltonian_eq} with DFT calculated energies. The former are represented by solid lines in Fig.~\ref{model_vs_DFT_E_YELMO} and show excellent agreement with our total energy DFT calculations.


\section{Numerical results}

We will focus on calculating the magnetic trends for the following three compounds, in order of increasing radius, \textit{R} = Lu, Er and Y for which the Shannon radii for the octahedrally coordinated 3+ ions are given in Table~\ref{tab:deltax_inhRMO}. We begin by extracting the parameters $A$ and $J_z$, defined in Eqs.~\eqref{Jz_av} and \eqref{A_av} at K\textsubscript{1} amplitudes $\delta x_{\text{Mn}}=-0.01,-0.005,0,0.005$ and $0.01$ which cover the spread of reported $\delta x_{\text{Mn}}$ values (Table \ref{tab:deltax_inhRMO}).

\subsection{Dependence of in-plane anisotropy on K\textsubscript{1} mode}

\begin{figure}[h!]
    \centering
\includegraphics[scale=0.295]{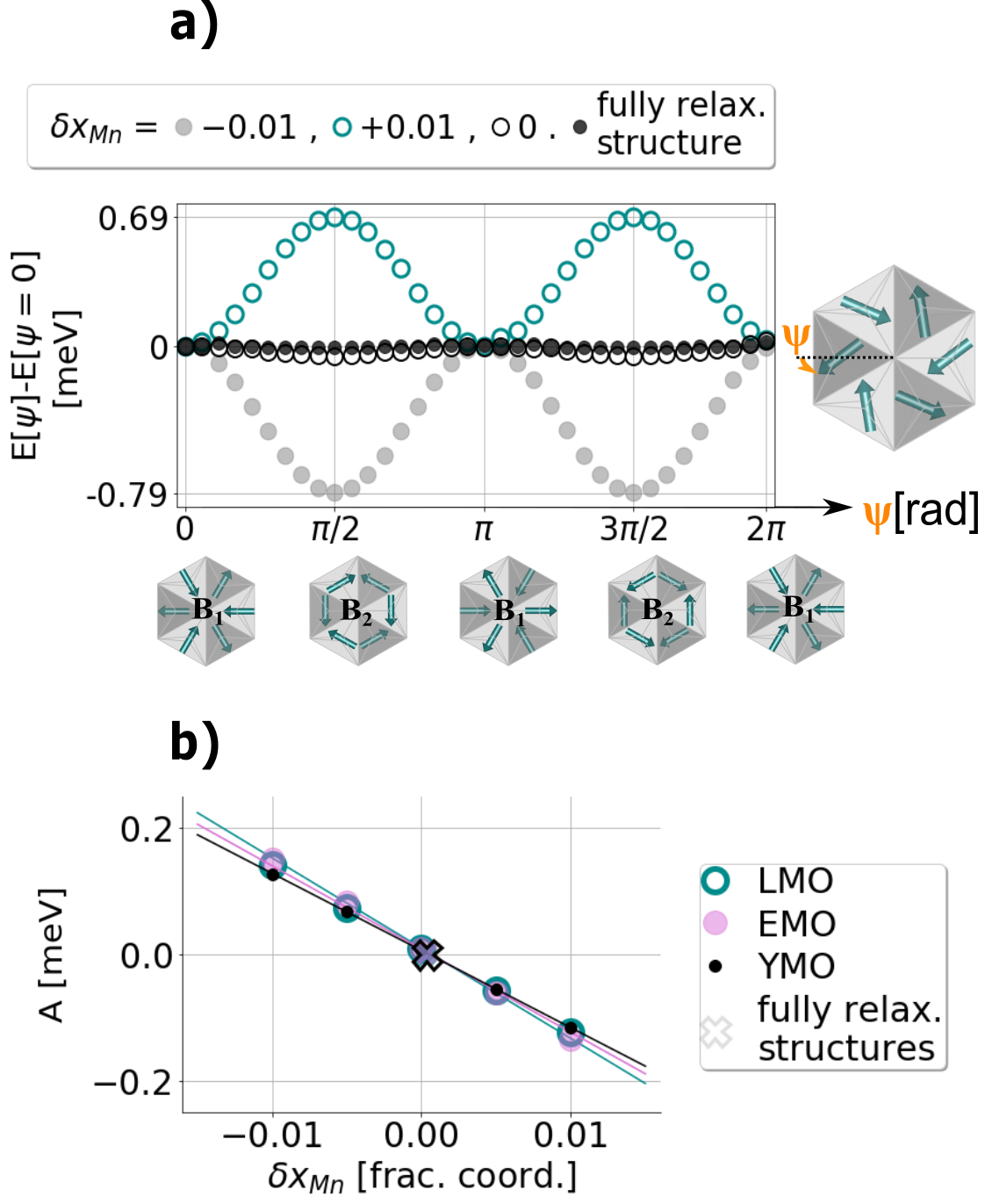}
    \caption{SIA energy landscape as a function of $\delta x_{\text{Mn}}$. \textbf{a)}~Total energy of YMnO\textsubscript{3} as a function of $\psi$  for three different K\textsubscript{1} geometries ($\delta x_{\text{Mn}}=-0.01,+0.01$ and 0) and for a fully relaxed (fully relax.) structure. $\psi$ describes the angle of an in-phase planar rotation of the \textbf{B\textsubscript{1}} configuration. The in-plane anisotropy $A$ is defined by the energy barrier at the local extrema. $A$ is negative (positive) for $\delta x_{\text{Mn}}=-0.01$ (+0.01) and is almost non-existent for $\delta x_{\text{Mn}}=0$.  \textbf{b)} $A$ calculated for \textit{R} = Lu,Y and Er for a range of $\delta x_{\text{Mn}}$ geometries (circles) and for their fully relaxed geometries (crosses). The full lines represent a linear least squares fit of the DFT calculated energies.}
    \label{SIAvsK1}
\end{figure}

Fig.~\ref{SIAvsK1}\textbf{a)} shows the calculated total energy change as a function of in-plane spin angle on four different geometries: a fully relaxed structure and three selective dynamics geometries corresponding to $\delta x_{\text{Mn}}=-0.01,0,0.01$. Spins located in two consecutive planes are rotated in-phase, so that only the energy contribution of $A$ is varied whilst keeping the $J_z$ exchange energy constant. Note that this rotation is different to the ones spanning the mK\textsubscript{2,3} modes in Fig.~\ref{SIAvsK1}, where both the in-plane anisotropy and inter-planar exchange energies change as a function of the rotation. We see that the angle dependence of the total energy E has opposite behaviour for $\delta x_{\text{Mn}}=-0.01$ and +0.01. E has its minimum value (-0.79\,meV) for B\textsubscript{2} order ($\psi=n\frac{\pi}{2}$) in the $\delta x_{\text{Mn}}=-0.01$ geometry, whereas the B\textsubscript{1} order ($\psi=n\pi$) minimises E (at 0.69\,meV) for $\delta x_{\text{Mn}}=+0.01$. This indicates that $A$, as defined by Eq.~\eqref{A2_eq}, has opposite values for opposite $\delta x_{\text{Mn}}$ displacements and shows that a linear combination of mK\textsubscript{1} and mK\textsubscript{2} magnetic symmetries is energetically unfavourable. A bigger in-plane anisotropy for $\delta x_{\text{Mn}}=-0.01$ than for $\delta x_{\text{Mn}}=+0.01$ can be attributed to a bigger K\textsubscript{1} mode displacement in the $\delta x_{\text{Mn}}=-0.01$ geometry (by $\simeq0.0035 \text{ }$\r{A}). Interestingly, the energy surface flattens for $\delta x_{\text{Mn}}=0$ but not completely as in the fully relaxed structure; the in-plane anisotropy has relatively small local minima at $\psi=n\frac{\pi}{2}$, like in the $\delta x_{\text{Mn}}=-0.01$ geometry. These local minima can again be attributed to the imperfect mapping between $\delta x_{\text{Mn}}$ and \textbf{K} mentioned in the Methods section.\par

We then calculate $A$ following \eqref{A_av} for all three compounds, as shown in Fig.~\ref{SIAvsK1}\textbf{b)}. Importantly, we see that the size of the single-ion anisotropy is similar in all three materials, and that the sign of $A$ changes at approximately K\textsubscript{1}=0 in each case. The sign change indicates a change in the preferred orientation of the spins within the easy plane.

\subsection{Dependence of inter-planar exchange on K\textsubscript{1} mode}

\begin{figure}[h!]
    \centering
\includegraphics[scale=0.3]{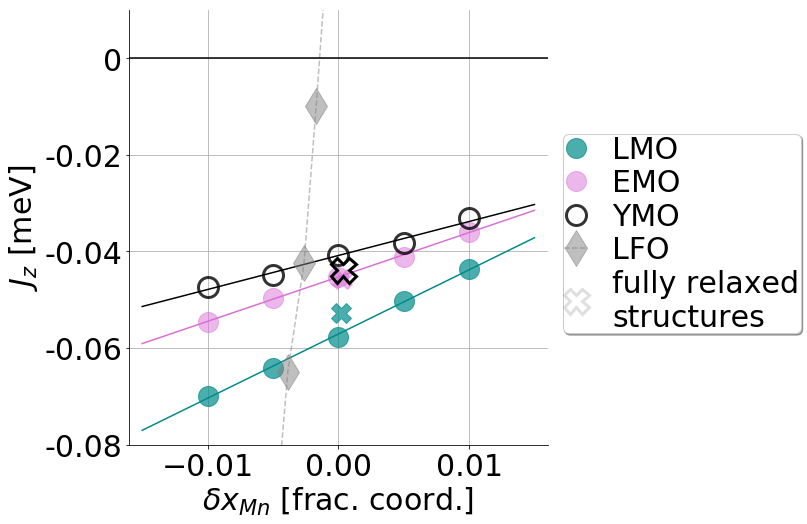}
    \caption{Circles and crosses represent $J_z$ values calculated in this work using Eq.~\eqref{Jz_av} on $\delta x_{\text{Mn}}\in[-0.01,0.01]$ and fully relaxed geometries, respectively . Grey diamonds connected by dotted lines show results for LuFeO\textsubscript{3} from Ref.~\cite{wang_structural_2014}.}
    \label{Jz_vs_K1}
\end{figure}

Next, we analyze the dependence of $J_z$ on the K\textsubscript{1} mode and show the calculated behaviour in Fig.~\ref{Jz_vs_K1}.  Over the range of K\textsubscript{1} values studied, $J_z$ is negative for all three materials. This corresponds to an FM (B-type) inter-planar interaction, with a linear dependence on K\textsubscript{1} consistent with the configurations observed in second harmonic generation measurements \cite{fiebig_determination_2000}. As the Mn ions shift away from their trimerisation centres, their apical oxygens shift by a smaller amount, leading to a decrease in the O\textsubscript{ap}-Mn orbital overlap and thus to a weaker inter-planar FM interaction. The displacement of the apical oxygens within the K\textsubscript{1} mode is similar across the three compounds, explaining the similarity in their $J_z[\textbf{K}]$ gradient. Additionally, the effective inter-planar exchange is stronger for systems with smaller radius \textit{R} cations and correspondingly bigger trimerisation, because $J_z$ reflects the difference between the two different inter-planar exchanges. This difference, illustrated in Fig.~\ref{symm_exch}, is larger for bigger bipyramidal tilts. Note that the inter-planar exchanges in the fully relaxed structures does not perfectly match the values for the $\delta_x{\text{Mn}}=0$ geometry. The difference could be due to the slight differences of atomic positions between the two structures mentioned in the Methods section.\par

For comparison, we show as grey diamonds in Fig.~\ref{Jz_vs_K1} the  $J_z$ values extracted for hexagonal LuFeO$_3$ by Wang et al. in Ref.~\cite{wang_structural_2014}. Hexagonal LuFeO\textsubscript{3} crystallises in the same structure as the hexagonal manganites and undergoes an analogous P6\textsubscript{3}/mmc to P6\textsubscript{3}cm structural phase transition. Its B-site chemistry differs from hexagonal manganites in that the Fe$^{3+}$ ions are in a formally $d^5$ high-spin state. As a consequence of this extra electron, LuFeO\textsubscript{3} has a stronger inter-planar interaction \cite{das_bulk_2014} and magnetic order sets in at a relatively higher Néel temperature ($\simeq150$\,K). The $J_z[\textbf{K}]$ trend of LuFeO\textsubscript{3} in Ref.~\cite{wang_structural_2014} is computed using the LDA+U method with U = 4.5\,eV and J = 0.95\,eV. Note that our studies using different choices of functionals (see Methods sections) suggest that this difference in behaviour is not a result of small differences in the choice of computational parameters. Fig.~\ref{Jz_vs_K1} clearly indicates that the stronger $J_{z}$ dependence on K\textsubscript{1} computed for LuFeO\textsubscript{3} \cite{wang_structural_2014} is necessary for small $\delta x_{\text{Mn}}$ distortions to favour the A-type ordering  observed in the hexagonal ferrites \cite{wang_structural_2014,das_bulk_2014}.

\subsection{\textit{R}-site dependence}

We note that the dependence of $J_z$ and $A$ on the size of the K\textsubscript{1} mode is sensitive to the \textit{R} site, with LuMnO$_3$ showing the strongest variation and YMnO$_3$ the smallest. For $ -0.015 < \delta x_{\text{Mn}} < +0.015$ the $A$ and $J_z$ parameters span $\left|A_{\text{max}}-A_{\text{min}}\right|=(0.43,0.41,0.39)$ and $\left|J_{\text{z,max}}-J_{\text{z,min}}\right|=(-0.04,-0.03,-0.02)$ for LuMnO$_3$,  ErMnO$_3$ and YMnO$_3$ respectively. This trend correlates with the size of the \textit{R}-site radius, the smallest radius (Lu$^{3+}$) having the largest polyhedral tilts \cite{kumagai_observation_2012} and in turn the largest change of the magnetic interactions with K\textsubscript{1}. This effect is captured by the first term in the Landau free energy expression \eqref{PK1K3_LFE} in which the K\textsubscript{1} mode amplitude $\mathcal{K}$ is linearly coupled to the square of the K\textsubscript{3} mode amplitude $\mathcal{Q}$, as well as by the coupling of $\mathcal{K}$ to the squares of the two magnetic order parameters $\mathcal{M}_{1,2}$ in Eq.~\eqref{K1M_LFE}.

\subsection{Ground state phase diagram}

\begin{figure}[h!]
\centering
\includegraphics[scale=0.37]{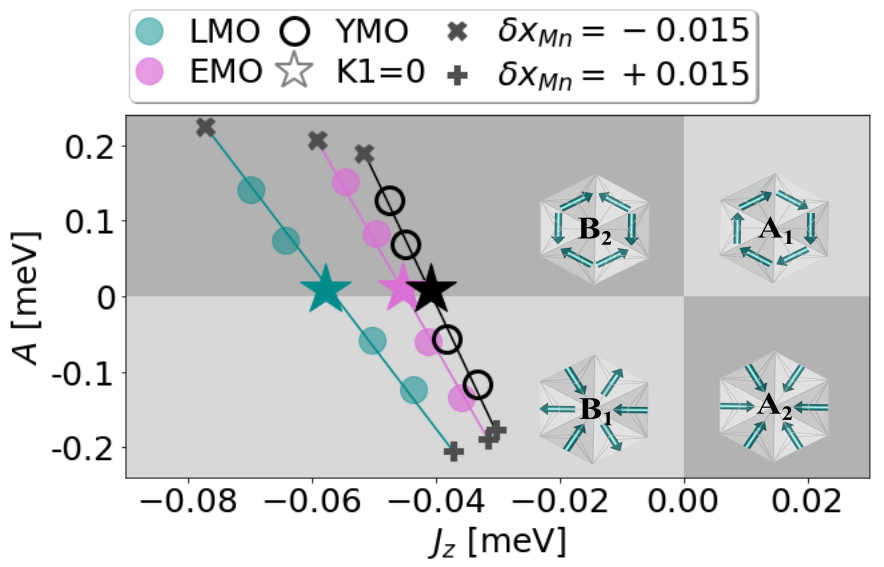}
\caption{Magnetic ground state phase diagram in the model Hamiltonian parameter phase space. Circles represent extracted $(J_z,A)$ pairs and full lines are linear least-square fits plotted for the range $\delta x_{\text{Mn}}\in[-0.015,0.015]$, going from top left to bottom right. Points marked by stars are the extracted values for $\delta x_{\text{Mn}}=0$. Crosses and plus markers indicate parameters for $\delta x_{\text{Mn}}=-0.015$ and +0.015, respectively.}
\label{fig:phase_diagram}
\end{figure}

We summarise our calculated parameters and their effect on the magnetic ground state  in the phase diagram of Fig.~\ref{fig:phase_diagram}, where we plot the calculated $A$ as a function of the calculated $J_{z}$ over the interval $\delta x_{\text{Mn}}\in[-0.015,0.015]$ for the three compounds. Since magnetic order is known unambiguously, this phase diagram allows us to predict the direction of the manganese displacement for which there is still a large spread in the experimental data. The fact that the B\textsubscript{1} state is measured in YMnO$_3$ and the B\textsubscript{2} state in ErMnO$_3$ implies opposite $\delta x_{\text{Mn}}$ displacements in the two materials.\par

\begin{table}[h]
\begin{tabular}{c c c}
    \midrule

    \textbf{Hamiltonian} \quad& \textbf{Landau} & \textbf{Ground state} 
         
    \vspace{0.25cm}\\
         
   \begin{tabular}{@{}c} $A>0$ \\ $J_z<0$\\ \end{tabular}  & 
   \begin{tabular}{@{}c} $|\alpha_3|>|\alpha_2|$ \\ $\alpha_3<0$ \\\end{tabular} &
   
   \begin{tabular}{@{}c} \textbf{B\textsubscript{2}} \\ ($\psi_3=n\pi$) \\\end{tabular}  \\
    \midrule
    
   \begin{tabular}{@{}c} $A<0$ \\ $J_z<0$\\ \end{tabular}  & 
   \begin{tabular}{@{}c} $|\alpha_2|>|\alpha_3|$ \\ $\alpha_2<0$ \\\end{tabular} &
   
   \begin{tabular}{@{}c} \textbf{B\textsubscript{1}} \\ ($\psi_2=n\pi$) \\\end{tabular}\\
    \midrule
   \begin{tabular}{@{}c} $A>0$ \\ $J_z>0$\\ \end{tabular}  & 
   \begin{tabular}{@{}c} $|\alpha_2|>|\alpha_3|$ \\ $\alpha_2>0$ \\\end{tabular} &
   
   \begin{tabular}{@{}c} \textbf{A\textsubscript{1}} \\ ($\psi_2=n\frac{\pi}{2}$) \\\end{tabular}\\
    \midrule
   \begin{tabular}{@{}c} $A<0$ \\ $J_z>0$\\ \end{tabular}  & 
   \begin{tabular}{@{}c} $|\alpha_3|>|\alpha_2|$ \\ $\alpha_3>0$ \\\end{tabular} &
   
   \begin{tabular}{@{}c} \textbf{A\textsubscript{2}} \\ ($\psi_3=n\frac{\pi}{2}$) \\\end{tabular}  \\
    \midrule

\end{tabular}
\caption{Mapping between Landau K\textsubscript{1}-mK\textsubscript{2,3} coupling parameters ($\alpha_2$ and $\alpha_3$), the model Hamiltonian parameters ($J_z,A$) and the magnetic ground state.}
\label{LFE_Ham_GS_summary}
\end{table} 

The phase diagram also sheds light on the origin of spin re-orientations observed in some hexagonal manganites.  Any change in the spin arrangement should correlate with the change in the Mn position as this creates transitions between A and B-type as well as between radial and tangential magnetic orders. For example, the B\textsubscript{2} to B\textsubscript{1} re-orientation in LuMnO\textsubscript{3} \cite{fiebig_determination_2000} as temperature decreases suggests a corresponding Mn displacement from negative to positive $\delta x_{\text{Mn}}$ values. Interestingly, hexagonal LuFeO\textsubscript{3} has a sub-T\textsubscript{N} spin reorientation, similarly to LuMnO$_3$, but from B\textsubscript{2} to A\textsubscript{2} as the temperature is lowered through 140\,K \cite{wang_structural_2014}. This translates to not only a change in the sign of $A$ (which is the case in LuMnO\textsubscript{3}) but additionally, to a change in the sign of the inter-planar exchange interaction. This crossing from the left (B-type order) to the right (A-type order) part of our phase diagram is attributed to the steeper $J_z[\textbf{K}]$ trend predicted for LuFeO\textsubscript{3} than for LuMnO\textsubscript{3} in Fig.~\ref{Jz_vs_K1}.\par

Finally, in Table~\ref{LFE_Ham_GS_summary} we summarise the mapping between the $A$ and $J_z$ magnetic Hamiltonian parameters to the K\textsubscript{1}-mK\textsubscript{2,3} coupling parameters $\alpha_2$ and $\alpha_3$ and to the magnetic ground state. For example, a radial arrangement of spins combined with an AFM inter-planar order (\textbf{A\textsubscript{2}}) corresponds to a ($A<0$, $J_z>0$) pair, which is equivalent to the following conditions on the K\textsubscript{1}-mK\textsubscript{2,3} coupling parameters: $|\alpha_3|>|\alpha_2|$ and $\alpha_3>0$. We have thus managed to described the magnetic ground state of hexagonal Manganites in two different spaces: the phase space of the Landau free energy parameters ($\alpha_2$,$\alpha_3$) as well as of the magnetic Hamiltonian parameters ($A,J_z$).


\section{Summary and Outlook}

In summary, this work rationalizes the observed evolution of the magnetic ground state across the hexagonal manganite series, the reported sub T\textsubscript{N} spin re-orientations in LuMnO$_3$, and explains why A-type magnetic configurations are energetically unfavourable  \cite{fiebig_determination_2000}.\par

To address the questions prompted by the experimental measurements of the Mn positions presented in Table \ref{tab:deltax_inhRMO}, we have determined, using symmetry arguments, the allowed coupling terms for $\delta x_{\text{Mn}}\neq0$ by extending earlier expansions of the Landau free energy \cite{fennie_ferroelectric_2005,das_bulk_2014, artyukhin_landau_2014}. The extended Landau expansion reveals a non-linear coupling of the K\textsubscript{1} mode to the K\textsubscript{3} and $\Gamma_2^-$ structural distortions, as well as to the magnetic order parameters. Furthermore, we have shown how $\delta x_{\text{Mn}}\neq0$ changes the magnetic interactions, using a model magnetic Hamiltonian with DFT-computed parameters, singling out the easy-plane anisotropy and the inter-planar symmetric exchange terms as the relevant magnetic interactions that lift the magnetic energy degeneracy between the A\textsubscript{1}, A\textsubscript{2}, B\textsubscript{1} and B\textsubscript{2} orders. These two findings support experimental measurements of sub-T\textsubscript{N} magnetoelasticity \cite{chatterji_magnetoelastic_2012,thomson_elastic_2014,lee_giant_2008} and non-zero, temperature-dependent $\delta x_{\text{Mn}}$ values, and suggest that the K\textsubscript{1} mode is responsible for both. In addition, our calculations indicate that, in materials with smaller \textit{R}-site radii (and correspondingly larger  K\textsubscript{3} distortions), the energy is more sensitive to changes in $\delta x_{\text{Mn}}$, motivating more precise experimental measurements of this quantity.\par

 Our work indicates that the K\textsubscript{1} mode provides a gateway to controlling the magnetic order in the h-\textit{R}MnO$_3$ series by varying its amplitude. We hope that this finding stimulates future studies investigating how the K\textsubscript{1} mode can be modified using external stimuli such as strain or doping. 


\section*{Acknowledgements}
This work was funded by the European
Research Council (ERC) under the European Union’s
Horizon 2020 research and innovation program project
HERO grant agreement No. 810451 and by the MARVEL national centre of competence in research (NCCR). Computational resources were provided by ETH Zürich and the Swiss National Supercomputing Centre, project IDs eth3 and s889. 

\newpage


%

\clearpage

\newpage

\onecolumngrid


\section*{Supplementary information} 

\subsection*{Effective $J_z$ and $A$ parameters extracted from total Energy calculations of the A\textsubscript{1,2} and B\textsubscript{1,2} magnetic configurations}
  \begin{table}[H]

      \centering
    \begin{tabular}{c|c|c|c|c|c|c|c|c}
        \multicolumn{2}{c}{}     & \multicolumn{3}{|c}{$\delta x_{\text{Mn}}$}  & & \multicolumn{3}{c}{$\delta x_{\text{Mn}}$}\\
    \cline{3-9}
        & $J_z$ & -0.01 & 0.0 & 0.01 & $A$ & 0.01 & 0.0 & 0.01 \\
    \hline
    
       \multirow{2}{*}{LMO} & (\ref{Jz1_eq}) &-0.06(7) &-0.05(5) &-0.04(1) &(\ref{A1_eq}) &0.14(0) &-0.00(2) &-0.14(6)\\
         & (\ref{Jz2_eq}) & -0.07(3)&-0.06(1) &-0.04(6) &(\ref{A2_eq}) &0.16(3) &0.02(2) &-0.12(4)\\
         \midrule
    \multirow{2}{*}{EMO} & (\ref{Jz1_eq}) &-0.05(4) &-0.04(4) &-0.03(4)&(\ref{A1_eq}) & 0.13(7)&0.00(2) &-0.13(0)\\
     & (\ref{Jz2_eq}) & -0.05(6)&-0.04(7) &-0.03(7) & (\ref{A2_eq})&0.14(4) &0.01(5) &-0.11(7)\\
        \midrule
     \multirow{2}{*}{YMO} & (\ref{Jz1_eq}) &-0.04(6) &-0.04(0) &-0.03(2)& (\ref{A1_eq})&0.12(2) &0.00(4) &-0.12(3) \\
                           & (\ref{Jz2_eq}) &-0.04(9) &-0.04(2) &-0.03(5)&(\ref{A2_eq}) &0.13(1) &0.01(2)& -0.01(1)\\
    \end{tabular}
    \caption{Extracted Hamiltonian parameters ($J_z$, $A$) following Eq.~\eqref{Jz1_eq} and Eq.~\eqref{Jz2_eq} for Lu,Er,YMnO\textsubscript{3}}
    \label{Jz_A_eff_TOTE}

\end{table}

The parameters used for the linear fit that ultimately produces Fig.~\ref{fig:phase_diagram} are the average of the two values obtained using equations Eq.~\eqref{Jz1_eq} and Eq.~\eqref{Jz2_eq} for $J_z[\delta x_{\text{Mn}}]$ and Eq.~\eqref{A1_eq} and Eq.~\eqref{A2_eq} for $A[\delta x_{\text{Mn}}]$ presented in table \ref{Jz_A_eff_TOTE}.

\subsection*{Four-state method extracted parameters}

\begin{figure}[H]
    \centering
    \includegraphics[scale=0.4]{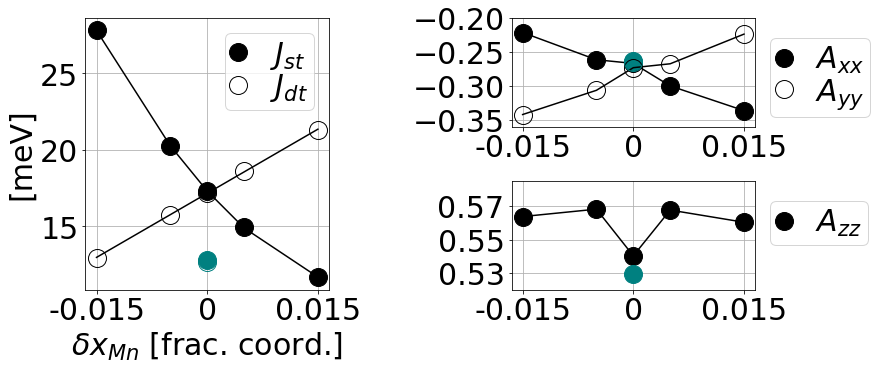}
    \caption{Four-state method extracted symmetric exchanges for first nearest neighbours belonging to the same trimer ($J_{st}$) and different trimer ($J_{dt}$) vs K\textsubscript{1} for YMnO\textsubscript{3}. Black symbols indicate parameters extracted for SD relaxed geometries, the corresponding parameters extracted on fully relaxed geometries are represented by the same symbols in teal. Energies (E) are calculated using the Ceperly and Alder local density approximation \cite{ceperley_ground_1980,lda_vasp}, Y\_sv, Mn\_sv and O pseudopotentials and U=4\,eV and J=1\,eV.}
    \label{4state para}
\end{figure}

Parameters were also extracted via the four-state method \cite{xiang_predicting_2011} for YMnO\textsubscript{3}. As expected, the first nearest neighbour interactions, $J_{st}$ and $J_{dt}$, are AFM and large, with values ranging from 10 - 30 meV, as shown in Fig.~\ref{4state para}. Their strength is a direct function of the Mn-Mn distance. A negative (positive) value of $\delta x_{\text{Mn}}$ corresponds to a contraction (expansion) of the triangle of Mn sites belonging to the same trimer which is reflected in the negative slope of $J_{st}$ in Fig.~\ref{4state para}. On the contrary, different trimer Mn sites are drawn farther apart (closer together) with negative (positive) $\delta x_{\text{Mn}}$; this behaviour is illustrated by the upwards slope of $J_{dt}$ in Fig.~\ref{4state para}. SIA in-plane coefficients match the effective $A$ values calculated via the method presented in the paper in Eq.~\eqref{A1_eq} and Eq.~\eqref{A2_eq}.

\end{document}